\documentclass{emulateapj}
\usepackage{graphicx,amsfonts,natbib}
\shorttitle{The Mass Spectra of GMCs}
\shortauthors{Rosolowsky}
\begin{document}
\title{The Mass Spectra of Giant Molecular Clouds in the Local Group}
\author{E. Rosolowsky\altaffilmark{1}}
\affil{Center for Astrophysics, 60 Garden St. MS-66, Cambridge, MA 02138}
\email{erosolow@cfa.harvard.edu}
\altaffiltext{1}{National Science Foundation (NSF) Astronomy and Astrophysics
Postdoctoral Fellow}
\begin{abstract}
We reanalyze the catalogs of molecular clouds in the Local Group to
determine the parameters of their mass distributions in a uniform
manner.  The analysis uses the error-in-variables method of parameter
estimation which accounts not only for the variance of the sample when
drawn from a parent distribution but also for errors in the mass
measurements.  Testing the method shows that it recovers the
underlying properties of cumulative mass distribution without bias
while accurately reflecting uncertainties in the parameters.  Clouds
in the inner disk of the Milky Way follow a truncated power-law
distribution with index $\gamma=-1.5\pm 0.1$ and maximum mass of
$10^{6.5}~M_{\odot}$.  The distributions of cloud mass for the outer
Milky Way and M33 show significantly steeper indices
($\gamma_{\mathrm{OMW}}=-2.1\pm 0.2$ and
$\gamma_{\mathrm{M33}}=-2.9\pm 0.4$) with no evidence of a cutoff.
The mass distribution of clouds in the Large Magellanic Cloud has a
marginally steeper distribution than the inner disk of the Milky Way
($\gamma=-1.7\pm 0.2$) and also shows evidence of a truncation with a
maximum mass of $10^{6.5}~M_{\odot}$.  The mass distributions of
molecular clouds vary dramatically across the Local Group, even after
accounting for the systematic errors that arise in comparing
heterogeneous data and catalogs.  These differences should be
accounted for in studies that aim to reproduce the molecular cloud
mass distributions or in studies that use the mass spectrum as a
parameter in a model.
\end{abstract}

\section{Introduction}
The mass distribution of molecular clouds is one of the primary
characteristics of the their population.  In the inner disk of the
Milky Way, the mass distribution follows a power law with $dN \propto
M^{\gamma} dM$, $\gamma \sim -1.5$.  More recent surveys of molecular
clouds throughout the Local Group find that the mass spectrum also
follows a power-law, but the indices are steeper than in inner Milky
Way \citep[e.g.~][]{eprb03,nanten}.  Indeed, the mass
spectrum may be the {\it only} feature of the molecular cloud
population that varies between systems, since other cloud properties
(e.g.~cloud radius and line width) obey the relationships
established in the Milky Way \citep{ws90,rpeb03,nanten}.  Careful
attention to accurately determining the parameters of the mass
spectrum is critical in using the mass spectrum to quantify
differences between cloud populations.  In addition, the empirically
derived mass distribution is an important parameter for theoretical
and modeling work.  Several studies aim to reproduce the mass
distribution of molecular clouds
\citep{kwan79,fractal-mspec,vs97,stutzki98,wada00} or use the mass spectrum
as inputs to models \citep{mw97,tan00,krumholz05}.  Most of these
studies focus on the canonical value of $\gamma\approx-1.5$ adopted
from the inner Milky Way, neglecting any variation in the
distribution.  Judging from the scope of these other studies,
measuring the mass distribution of molecular clouds is essential for
understanding both cloud formation and the importance of star-forming
clouds in regulating large scale star formation.

Since the parameters of the cloud mass distribution are widely used in
the study of the star-forming interstellar medium, this paper outlines
some of the pitfalls associated with the standard methods of
estimating the parameters of power-law distributions and suggests
improvements to minimize inaccuracy (\S\ref{fitting}).  With these
improvements, we reanalyze data from existing catalogs of molecular
clouds (\S\ref{datasec}) and note interesting results
(\S\ref{discussion}).  This work stresses the importance of accounting
for the observational uncertainties and systematic effects that
bedevil the study of molecular clouds.  Accurately deriving the index
of a power-law distribution is also useful for studying populations of
other objects.  In particular, the derived mass spectrum of clumps
within molecular clouds is subject to identical systematics as the
mass distributions studied in this work.  The methods developed in
this study as well as their attendant cautions are directly
applicable to the study of clump mass distributions and their
relevance in the formation on individual stars
\citep[e.g.~][]{clumpfind,gaussclumps}.  Luminosity and mass
distributions of stars and galaxies are characterized by non-linear
distributions and the techniques presented in this paper readily
extend to the study of these objects.

\section{Fitting Mass Spectra}
\label{fitting}
The mass distribution of a population of molecular clouds is usually
expressed in differential form, namely the number of clouds that would
be found in a range of masses.  In the limit of a small mass bin, this
is expressed as
\begin{equation}
\label{diff}
\frac{dN}{dM} = f(M).
\end{equation}
This expression can be integrated to give the cumulative mass
distribution 
\begin{equation}
\label{cum}
N(M' > M) = \int_{M_{max}}^{M'} f(M) dM = g(M),
\end{equation}
which gives the number of clouds with masses greater than a reference
mass as a function of that reference mass.  For molecular clouds, both
forms of the mass spectrum obey power-laws: $f(M)\propto M^{\gamma}$
and $g(M)\propto M^{\gamma+1}$ with $\gamma<-1$ in all known cases.
Some mass distributions lack clouds above some maximum mass $M_0$.  To
account for this feature, we adopt a truncated power-law distribution
as suggested by \citet{wm97} and alter their formalism to our
notation.  The full form of the cumulative distribution is
\begin{equation}
N(M'>M)= N_0 \left[\left(\frac{M}{M_0}\right)^{\gamma+1}-1\right],
\label{cumdist}
\end{equation}
where $M_0$ the maximum mass in the distribution. $N_0$ is the number
of clouds more massive than $2^{1/(\gamma+1)}M_0$, the point where the
distribution shows a significant deviation from a power law.  If
$N_{0} \sim 1$, there is no such deviation.  For this
form of the cumulative mass distribution,
\begin{equation}
\frac{dN}{dM}=(\gamma+1)\frac{N_0}{M_0}\left(\frac{M}{M_0}\right)^{\gamma},~
M<M_0.
\end{equation}
In most studies, only the index $\gamma$ is reported since $N_0$ is
assumed to be 1 and $M_0$ is the maximum mass cloud in the sample.
The index is the most important parameter since it describes how the
integrated mass is distributed between the high and low mass members
of the cloud population.  For values of $\gamma > -2$, the majority of
the mass is contained in the high mass clouds and the reverse is true
for $\gamma < -2$.  When $\gamma < -2$, the integrated mass diverges
as $M\to 0$, implying a break in the power-law behavior of the mass
spectrum at or below the completeness limit to ensure a finite
integrated mass.  Distributions with $N_0>1$ are also physically
interesting since they have a characteristic feature in an otherwise
featureless mass distribution.  In the Milky Way, \citet{wm97} report
evidence that $N_0$ is significantly different from unity, implying a
cutoff at high mass ($3\times 10^6~M_{\odot}$) in the Galaxy.  The
parameters of the mass distribution are important both as predictions
of theories as well as inputs to models. It is critical to estimate
these parameters with minimum bias from the mass measurements of a
cloud population.

\subsection{Binned Mass Spectra}
Most studies of the mass spectrum of giant molecular clouds (GMCs)
estimate the slope of the mass spectrum by fitting an approximation of
the differential version.  They generate this approximation by
separating the mass measurements into logarithmically spaced bins.
Then, the number in each bin ($N_{bin}$) is divided by the width of
the bin $\Delta M$: $dN/dM \approx N_{bin}/\Delta M$.  The
uncertainties in these bins are then assumed to arise from counting
errors, so $\sigma_{bin} = \sqrt{N_{bin}}/\Delta M$.  Studies using
this technique include \citet{srby87,wm97,eprb03,nanten,hc01} and many
others.

\begin{figure*}
\begin{center}
\plottwo{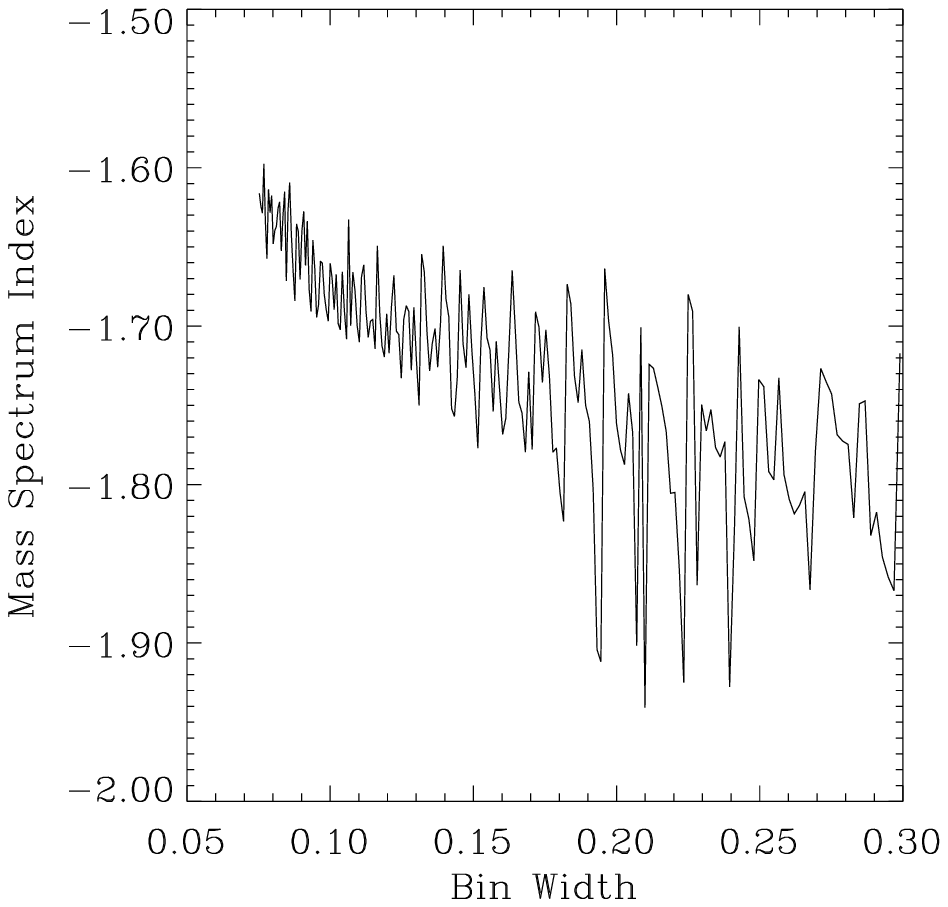}{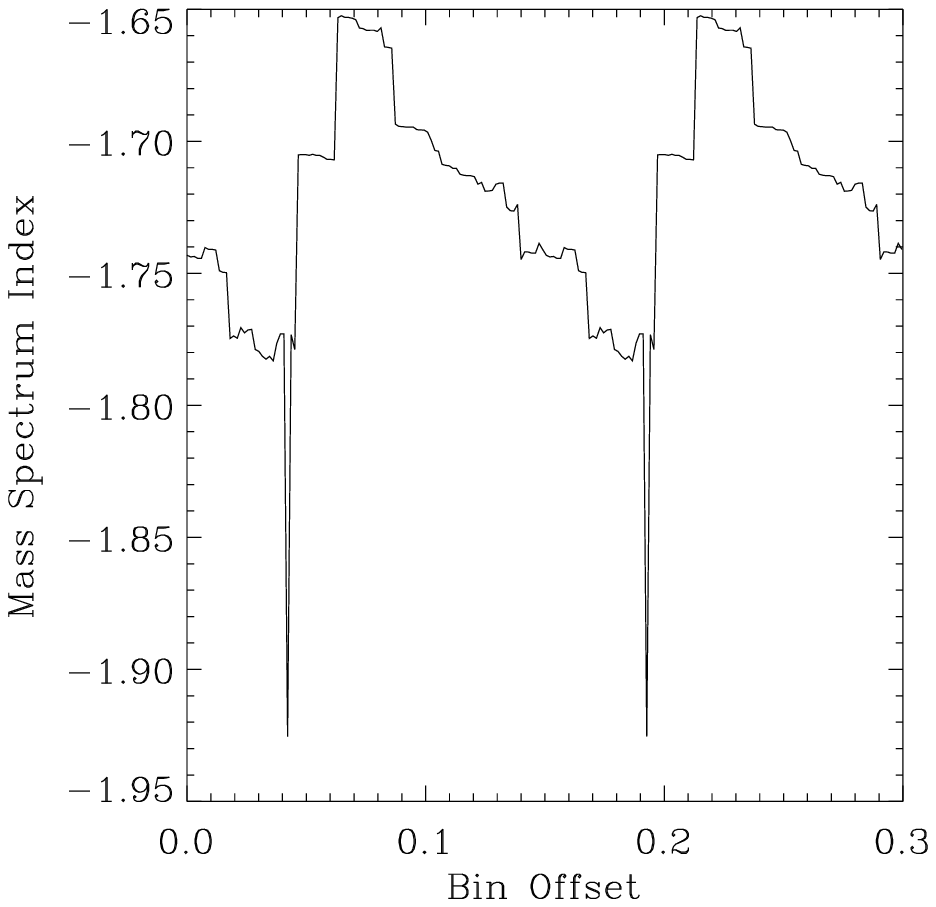}
\caption{Demonstration that fitting binned mass spectra is sensitive to
choice of bin size and offset.  The left panel shows the derived
index of the mass spectrum for the clouds in \citet{srby87} as a
function of the width of the logarithmic bins (in dex) used in
fitting.  The fits used the same completeness limit as the original
study ($7\times 10^{4} M_{\odot}$).  The bin size ranged from a quarter
octave $0.25\log_{10}(2)$, to a full octave, $\log_{10}(2)$.  The
right panel shows the variation in mass spectrum for half octave bins
and different bin positions.  The bins are shifted in log-space by the
quoted bin offset.  Both variations in parameters show significant
variation in the derived index. \label{binsuck}}
\end{center}
\end{figure*}

There are two principal drawbacks to this technique: (1) it is
sensitive to the selected values of bin size and bin spacing and (2)
it neglects errors in the mass determination of the clouds, which can
be substantial. Figure \ref{binsuck} shows the variation in the
derived index of the mass spectrum for different choices of bin size
and bin position.  To generate these figures, we used the mass data
from \citet[][SRBY]{srby87} with the same completeness limit of
$7\times 10^{4} M_{\sun}$ as is quoted in their paper.  For a given
set of bin parameters, we fit a power-law differential mass spectrum
to the results to all data that are at least one full bin above the
completeness limit.  We follow the method of \citet{wm97} for the fit
and the determination of errors in the mass distribution.  The
systematic error in the parameters is comparable the errors typically
quoted in these studies.  Such errors become negligible in the limit
of large numbers of clouds.  In the study of
\citet[][HCS]{hc01}, there are over 1300 clouds above the completeness
limit as opposed to only 200 in the SRBY study.  When the same
experiment is performed on this much larger sample, the variation in
the derived index reduces to $\pm 0.05$ and agrees with the $-1.8$
quoted in the HCS paper.  To use binned mass spectra in estimating the
parameters of the mass distribution, the sample should have $N_{clouds}
> 500$ to reduce errors to less than 0.1 in the index.

In addition to large variations in derived bin parameters, the binning
method also neglects the principal source of uncertainty, namely the
mass measurement itself.  The mass of a molecular cloud is notoriously
difficult to calculate.  The principal methods for deriving the mass
are using the CO-to-H$_2$ conversion factor and using the virial
theorem.  The conversion factor linearly scales the integrated CO
surface brightness to a column density along a line of sight.  With a
distance measurement, the column density of the cloud and the area on
the sky are combined to calculate the cloud mass.  The conversion
factor is empirically tested to trace H$_2$ column density across a
variety of environments \citep{bloemen} though variation is reported
among galaxies \citep{xfac96}.  Within a single galaxy, however, the
conversion factor has been found to be constant despite changes in the
galactic environment \citep{rpeb03}.  The virial mass measurement
assumes that clouds are virialized and uses the resolved cloud sizes
($R_e$) and line widths ($\sigma_v$) to convert to a virial mass:
$M_{\mathrm{VT}} = 5 R_e \sigma_v^2/(\alpha G)$ where $\alpha$ is the
virial parameter, which depends on the mass distribution within the
cloud as well as the influence of magnetic fields and external
pressure on the energy balance of the cloud.  HCS present evidence
that molecular clouds with $M<10^4~M_{\odot}$ are not virialized and
the virial mass measurement overestimates the masses of these clouds.

Both of these methods for measuring cloud masses are subject to large
($\lesssim 50\%$) errors.  Absolute flux calibration of CO data is
rarely accurate to better than 10\%.  The variations of the conversion
factor with physical conditions remain poorly understood, despite many
attempts to quantify them \citep{xfac-wolfire,xfac-dickman}.  Finally,
the distances to most molecular clouds are difficult to measure.  For
Milky Way molecular clouds, most distances are determined
kinematically with the distance degeneracy for the inner Galaxy being
broken by angular scale, displacement above the plane, and association
with other objects of known distance (SRBY).  Distance measurements
are also important in measuring virial masses since the physical size
of the cloud is determined by converting an angular scale to a
physical length.  In addition, small or distant clouds are often
poorly resolved and great care must be taken to measure the radius of
an intensity distribution that has been convolved with the telescope's
beam.  The largest pitfall in the virial method is the question of its
applicability.  Mass measurements nearly always neglect other
contributions that are present in the full virial theorem, such as
external pressure, changing moment of inertia, magnetic fields, the
degree of virialization and the measurement of a single size for a
triaxial system.  These deviations are frequently parameterized using
the virial parameter (see above) which is surprisingly constant for
massive molecular clouds ($\alpha \sim 1.5$, SRBY, HCS).  Thus, the
virial mass estimate provides a reasonable measurement of a cloud's
dynamical mass.  With all these potential sources of error, the
masses of molecular clouds are highly uncertain, often to 50\%, and
this uncertainty should be included in the determination of the mass
spectrum parameters.

\subsection{Cumulative Mass Spectra}
When a sample contains only a small number of clouds ($N_{clouds} <
500$), it is still possible to derive the parameters of a mass
spectrum by fitting the cumulative distribution of masses.  Recent
work by \citet{nanten-mspec} demonstrated the utility of this method
for clouds in the LMC.  The principal difficulty in using this method
arises in assessing errors to the data in the cumulative mass
spectrum. Uncertainties appear both in the mass of the cloud and in
the variance of a random sample being drawn from an infinite parent
distribution.  Practically, this results in fitting a truncated
power-law function to data with errors in both coordinates.  The mass
coordinate has an uncertainty from the measurement error and the
cumulative number has an uncertainty characterized by a counting
error, equal to $\sqrt{N}$.

To fit the data, we use the ``error-in-variables'' method for
parameter estimation in non-linear functions that have uncertainties
in both coordinates.  The method was developed by \citet{errinvar}
which, in turn, is the full development of a method originally
suggested by \citet{deming}.  An equivalent method was developed into
an algorithm by \citet{errinvar-alg} which has been incorporated into
StatLib\footnote{\url{http://lib.stat.cmu.edu/}}.  It is this
algorithm upon which the present work is based although the
error-in-variables method was also presented to the astronomical
community with the work of \citet{jefferys1}. The method maximizes the
likelihood that a set of data $(M,N)$ with associated uncertainties
$(\sigma_M,\sigma_N)$ can be drawn from a distribution with parameters
$\{N_0,M_0,\gamma\}$.  Since the equations of condition cannot be
solved algebraically for the parameters (as they can in the linear
case), the minimization is performed iteratively in two interleaved
phases.  First, the true values of the data ({\it i.e.}~without
measurement errors) are estimated by maximizing the likelihood of
being drawn from a distribution with some initial guess of parameters.
Then, using the estimate of the true values of the data, the optimal
values of the parameters are determined.  The process is iterated
until estimates of both the true data values and the parameters are
determined. 

Instead of performing the fit to the data with the model given by
equation \ref{cumdist}, we use the algebraically equivalent expression
\begin{equation}
y_i = \theta_1 x_i^{\theta_2}+\theta_3
\end{equation}
to improve independent estimates of $M_0$ and $N_0$ which are highly
covariant in the original formulation.  Once the algorithm has
converged on a vector of parameters, $\mathbf{\theta}$, we transform
the elements of $\mathbf{\theta}$ back to the parameters of interest.
We use a bootstrapping technique to estimate uncertainties in the
derived parameters, using 100 trials to sample the distribution of
derived parameters which is often non-Gaussian.  The quoted values of
the uncertainties in the parameters are the median absolute deviations
of the transformed parameter distribution from the bootstrapping
trials.  Examination of the parameter distributions using a large
number of bootstrap trials shows that the medians adequately
characterize the uncertainties. For some distributions, there are more
high-mass clouds than expected from the distribution at lower mass
(i.e.~the opposite of a truncation).  In this case, the parameter
$\theta_3$ converges to zero.  When this occurs, we fit a power law to
the distribution of form
\begin{equation}
N(M'>M)= \left(\frac{M}{M_0}\right)^{\gamma+1}
\label{purepl}
\end{equation}
and report only $M_0$ and $\gamma$.

We validated our method by fitting the model to random data drawn from
power-law distributions with known parameters.  The trial data have
normal deviates of known dispersion added to them that simulate the
effects of measurement error.  In these simulations, we find that the
method both recovers the properties of the distribution without bias
and produces error estimates from bootstrapping that agree well with
the scatter in derived parameters around the known parameters.  This
implies that we are properly accounting for the error in the sample as
well as recovering the properties of the underlying distribution.
These tests demonstrate that the error-in-variables fit to the
cumulative mass distribution should be favored over a fit to the
binned mass distribution.

\subsection{Systematic Effects}
In addition to the errors in the mass measurement, there are also
systematic errors in the generation of mass spectra.  The two dominant
contributions to the systematic errors are choice of the mass
measurement (virial vs.~luminous) and the method used to generate the
cloud catalog.  SRBY report $M_{VT}\propto M_{LUM}^{0.8}$ in their
sample which implies that determinations of the index $\gamma$ can
vary by $10\%$ depending on the mass measurement.

The process used to generate the catalogs is likely the dominant
systematic in measuring the parameters of the mass distribution.  In
particular, the resulting parameters of mass distributions depend on
the algorithm which assigns flux into the physically significant
substructures for which the masses are determined.  Such
decompositions include (1) human assignment into clouds
\citep[{e.g.~}][]{ws90} (2) assignment by grouping neighboring pixels
above a cut in brightness (SRBY, HCS) and (3) computer algorithms such
as CLUMPFIND \citep{clumpfind} or GAUSSCLUMPS \citep{gaussclumps}.
Assigning multiple distinct structures into a single cloud
artificially drives the index of the mass spectrum towards more
positive values.  Such blending is most likely to occur when using
kinematic data to detangle emission in the inner Milky Way.
Conversely, overzealous decomposition of objects can erroneously split
high mass objects into lower mass objects, decreasing the value of the
index or creating an artificial truncation in the distribution.
Predicting the quantitative impact of these systematics is beyond the
scope of this work.

Ideally, the mass distribution should be derived using the same
decomposition algorithms and mass determinations from both
observations and simulations to minimize these systematic effects.
However, the magnitude of these systematic effects can be estimated by
analyzing the same data set with different methods.  Using the derived
parameters from the heterogeneous data sets in this study, we find
that the index can be robustly determined, in spite of these
systematic effects.  Identifying truncations and maximum masses are
complicated by these systematic effects and require care to accurately
recover (see below).

\section{Local Group Mass Spectra}
\label{datasec}
\begin{figure}
\begin{center}
\plotone{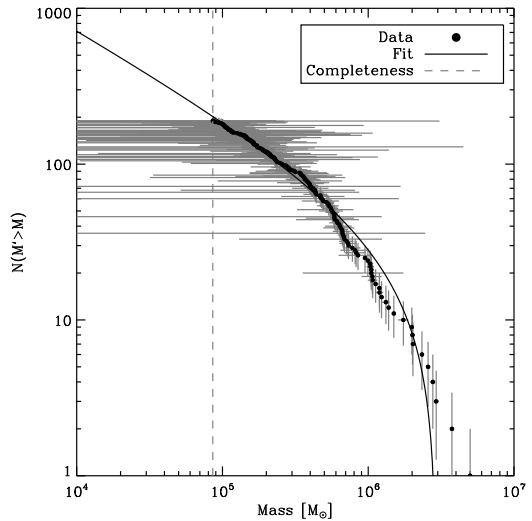}
\caption{ The mass distribution of the SRBY virial mass measurements.
A truncated power-law fit to the data using the methods of
this study is shown as a solid line.  The data show a significant
break around $N=50$ and the fit recovers this feature well. 
\label{srby-example}}
\end{center}
\end{figure}

Using fits to the cumulative mass distribution, we have reanalyzed the
catalogs of GMCs in the Local Group.  Our results show significant
differences in the mass distributions of the GMC populations.  For
each of the catalogs discussed below, we fit a power-law to the
cumulative mass distribution, including a truncation if appropriate
(see above).  Unless otherwise stated, when both virial and luminous
measurements of the mass are reported, we use an error equal to half
the difference between the two mass measurements plus a 10\% flux
calibration error added in quadrature.  The results of the new fits to
the Local Group mass distributions are summarized in Table \ref{data}.
The reported errors are the median absolute deviation of the derived
parameters for 100 bootstrapping trials.  To illustrate a fit to the
data, we plot the results of the fit to the virial mass data of
\citet[][SRBY]{srby87} in Figure \ref{srby-example}.

\begin{deluxetable*}{ccccccc}
\tablecaption{\label{data}Parameters of Mass Spectra for GMCs 
in Local Group Studies}
\tablewidth{0pt}
\tablehead{
\colhead{Object} & \colhead{Name} & \colhead{Type} & 
\colhead{Num.} & \colhead{$\gamma$} & \colhead{$N_0$} & \colhead{$M_0/(10^5~M_\odot)$}}
\startdata
Inner MW & SRBY & VT &  190 & $-1.53 \pm 0.07$ & $36. \pm 12.$ & $29. \pm 5.0$ \\
Inner MW & SRBY & CO &  173 & $-1.53 \pm 0.06$ & $27. \pm 11.$ & $41. \pm 9.5$ \\
Inner MW & SYSCW & VT &  107 & $-1.58 \pm 0.15$ & $14. \pm 10.$ & $26. \pm 7.6$ \\
Inner MW & SYSCW & CO &  97 & $-1.41 \pm 0.12$ & $21. \pm 13.$ & $29. \pm 7.2$ \\
Outer MW\tablenotemark{a} & HCS & VT &  227 & $-2.56 \pm 0.11$ & \nodata & $3.2 \pm 0.78$ \\
Outer MW\tablenotemark{a} & HCS & CO &  81 & $-2.06 \pm 0.15$ & \nodata & $6.3 \pm 3.1$ \\
Outer MW & BKP & VT &  336 & $-2.29 \pm 0.08$ & $4.5 \pm 3.5$ & $2.9 \pm 1.0$ \\
Outer MW & BKP & CO &  81 & $-2.16 \pm 0.17$ & $2.7 \pm 2.9$ & $2.0 \pm 1.0$ \\
M33 & EPRB & CO &  58 & $-2.85 \pm 0.36$ & $2.5 \pm 2.7$ & $8.6 \pm 3.3$ \\
LMC & NANTEN & VT &  44 & $-1.71 \pm 0.19$ & $10. \pm 6.5$ & $23. \pm 4.6$ \\
LMC & NANTEN & CO &  55 & $-1.72 \pm 0.12$ & $6.1 \pm 3.6$ & $82. \pm 32.$ \\
\enddata
\tablenotetext{a}{The mass distribution shows an excess of clouds at
high mass, implying there is no truncation in the sample.  A pure
power law has been fit to the data (Equation \ref{purepl}).}
\end{deluxetable*}

\subsection{The Inner Milky Way}
There are two major studies of GMCs in the inner Milky Way.  Both SRBY
and \citet[][SYSCW]{syscw} analyzed FCRAO survey data from the first
quadrant of the Galaxy decomposing the emission into clouds using
different algorithms.  Comparing the results of these two studies
highlights the systematic effects of using different decomposition
algorithms.  Both studies identified clouds as contiguous regions
above a fixed antenna temperature cutoff but chose different
thresholds and methods for decomposing substructure.  We use their
measurements for virial mass and luminous mass, correcting for
differences in virial definitions and galactic scales as summarized in
\citet{wm97}.  The index of the power-law is unaffected by the choice
of conversion factor.  Typical mass errors are factors of $\sim 15$\%.
We fit all clouds with masses greater than $1 \times 10^5~M_{\odot}$
in the SRBY study, which approximates their reported completeness
level.  For the SYSCW study, we compared the virial and luminous mass
measurements after scaling the data and we find that the virial mass
estimates are a factor of 2 higher than the luminous mass measurements
for the high mass clouds.  To place the samples of equal footing, we
scaled the luminous mass of the clouds by a factor of two to bring the
mass estimates into agreement.  We then examined the distributions and
established a completeness limit of $5\times 10^4~M_{\odot}$ based on
where the distribution departed from a power law on the low mass end.
Fitting to both the virial and luminous masses for both studies finds
an index $\gamma\approx -1.5$ and a significant cutoff with $\sim 25$
clouds at the cutoff.  For all four fits, $M_0\approx 3\times
10^6~M_{\odot}$.  The relatively small differences between the derived
parameters despite the different catalog methods suggests that
systematic effects are small in this case.  Since the cataloging
methods are conceptually similar in the two studies, this result is
not surprising.  The derived value of $N_0$ is slightly higher in the
SRBY method than SYSCW suggesting there is some influence of the
catalog method on the cutoff values.

The mass distribution for the inner Milky Way is shallower than found
for other systems.  Two effects may bias the results to a shallower
index.  First, line-of-sight blending will make several less massive
clouds appear as a single, more massive cloud shifting the index to
shallower values.  The methods used to generate the SRBY and SYSCW
catalogs do little to split up blends of emission.  Second,
incorrectly resolving the distance ambiguity will also bias the mass
distribution to shallow indices.  If every cloud has the same
probability of having its distance incorrectly determined, then more
low-mass clouds at the near distance will be erroneously counted as a
high-mass clouds at the far distance than the reverse, simply because
there are more low-mass clouds.  This latter bias can increase the
index of the mass distribution by as much as 0.2 for 20\% of clouds
being assigned to the wrong distance.  Thus, the index of the mass
distribution for the Inner Milky Way very likely is steeper than can
be derived from the current observational data.

\subsection{Outer Milky Way}
The data used for the Outer Milky Way are from the FCRAO survey of a
section of the second quadrant \citep{fcrao-ogs} which were
subsequently analyzed by both HCS and \citet[][BKP]{outercat}. HCS
used a cloud extraction algorithm similar to SRBY, but defined cloud
properties from the intensity distributions slightly differently.  In
contrast, BKP used a modified CLUMPFIND algorithm to identify peaks in
the emission distribution as the nuclei of distinct clouds.  Their
algorithm extracts roughly $\sim 50$\% more sources than the work of
HCS.  They assign cloud properties to the emission distribution in a
similar fashion as HCS.  Mass errors in the HCS study are given as
half the difference between mass measurements plus a flux error and
errors in BKP are reported in their study.  Since clouds in the outer
Galaxy with masses smaller than $10^4~M_{\odot}$ are not virialized,
we set $10^4~M_{\odot}$ as the lower mass limit for the fits to these
catalogs.  Adopting this truncation includes many more virial mass
measurements than luminous mass measurements since the virial mass
tends to overestimate the mass of clouds with
$M_{\mathrm{LUM}}<10^4~M_{\odot}$.  Thus, the luminous mass
distribution likely represents the underlying mass distribution better
than the virial mass distribution.  We also require the kinematic
distance to be larger than 2 kpc to minimize errors in the distance
determination.  We find that the index of the mass distribution is
steeper than reported in HCS, which is due to the improved fitting
methods ($\gamma =-2.1\mbox{ vs.~}-1.8 \mbox{ in HCS}$).  Since the
luminous mass is likely a better tracer of cloud mass, we also perform
a fit to the luminous mass data alone using a lower limit of $2\times
10^3~M_{\odot}$ and derive an index of $\gamma=-2.05\pm 0.06$.

The catalog of HCS shows more clouds than would be expected at high
mass given a power-law extrapolation from lower masses.  Such an
excess is not seen in the BKP catalog, because of the more aggressive
decomposition algorithm employed in the latter study.  Without careful
analysis of the individual clouds, it is impossible to say what
represents the true distribution of clouds at high mass in the outer
Galaxy. Since evidence for a cutoff appears in the BKP catalog but not
in the HCS catalog, comparing these two studies illustrates the
systematic effects of different catalog methods. There is not the
strong evidence for a truncation in the outer Galaxy that is found for
the inner Galaxy data.  This is likely because there are too few
molecular clouds to populate the distribution up to the truncation
mass.  Nonetheless, the index of the mass distribution is
well-determined and is significantly steeper than that found in the
inner Milky Way.

\subsection{M33}
M33 is the only spiral galaxy for which a catalog of GMCs exists with
a known completeness limit \citep[][EPRB]{eprb03}.  Since the galaxy
is seen from an external perspective, blending effects are
dramatically reduced compared to Milky Way studies.  However, there
are only 59 clouds above the reported completeness limit of $1.5\times
10^5 M_{\odot}$, and the clouds have only CO masses reported since the
individual clouds are not resolved.  A follow-up study \citep{rpeb03}
shows that the virial mass is proportional to the luminous mass for
GMCs in M33 and that the $M_{\mathrm{CO}}/M_{\mathrm{VT}}$ does not
vary significantly over the galaxy.  We estimate the error in their
measurements as the difference between the measured and corrected mass
discussed in EPRB plus their quoted 25\% calibration error in the flux
scale of the interferometer.  The derived value of the mass index
($\gamma = -2.9$) is very steep.  M33 is also the most distant galaxy
in this reanalysis and observational biases may affect the index of the
mass distribution.  However, the potential biases would only make the
mass spectrum appear shallower than it actually is.  In particular,
blending effects will make several less massive clouds appear as a
single massive cloud, and underestimates of the completeness limit will
cause the number of low-mass clouds to be underestimated.  The
influence of either of these effects would imply that the mass index
is actually steeper than what is measured: $\gamma \le -2.9$.

It is likely that the extremely steep slope of the mass distribution
is actually the tail of a distribution with a cutoff mass below the
completeness limit of the survey.  EPRB estimate a characteristic mass
between $3-7\times 10^{4}~M_{\odot}$, which could simply be a cutoff
mass in a truncated power-law distribution.  To illustrate the effects
of fitting a truncated power-law distribution above the cutoff mass,
we repeated the analysis of clouds in the inner Milky Way restricting
the sample to clouds near the cutoff mass ($M>2\times
10^6~M_{\odot}$).  Fitting to the restricted sample gives
$\gamma=-2.2$ with no evidence of truncation instead of $\gamma=-1.5$
with a truncation.  This supports our conjecture that the steep slope
of the M33 mass distribution can be attributed to fitting a power-law
distribution above the mass cutoff.

\subsection{Large Magellanic Cloud}
The only other complete survey of GMCs in a galaxy was completed using
the NANTEN 4-m telescope to observe the LMC.  \citet{nanten} report
the most recent catalog of GMCs, including 55 resolved GMCs for which
virial masses can be measured.  A subsequent paper
\citep{nanten-mspec} reports an index of $\gamma=-1.9$ using CO and
virial masses from a currently unavailable catalog of more GMCs.  All
of the resolved clouds have masses above the completeness limit of the
survey.  Using the virial masses for the 55 reported clouds, we derive
a mass spectrum index consistent with \citet{nanten-mspec} with some
evidence of truncation.  The index on the mass distribution derived
from the virial masses is likely a lower limit (i.e.~$\gamma > -1.9$)
because the reported virial mass measurements do not account for beam
convolution.  The error-in-variables fit to the data finds that the
mass distribution is shallower ($\gamma=-1.7\pm 0.2$) than reported in
\citet{nanten-mspec} with some evidence of a cutoff.
The maximum mass in the LMC is similar to that in the inner Milky Way
($3\times 10^6~M_{\odot}$), but the value is poorly constrained by the
limited number of clouds in the catalog.

\section{Discussion}
\label{discussion}
There is a real variation in the mass distribution of GMCs across the
Local Group with indices ranging from $\gamma=-2.9$ to $-1.5$. There
are cutoffs at a maximum mass of $10^{6.5}~M_{\odot}$ in catalogs from
the inner Milky Way and the LMC.  In general, the differences in the
mass distributions have been unappreciated or trivialized; but they
are, in fact, significant.  In the inner Milky Way, the top-heavy mass
distribution means that studying the most massive clouds encompasses
most of the star-formation in that part of the Galaxy.  In contrast,
low mass clouds contain a substantial fraction of the molecular mass
in the outer Milky Way and M33.  In systems with bottom-heavy mass
distributions, the star-forming properties of these low mass clouds
must be examined to obtain a complete picture of the star-forming ISM.
Using $\gamma\approx -1.5$ is appropriate for the inner Milky Way but
not for all galaxies.

Since molecular clouds {\it of a given mass} appear to be similar
across the Local Group
\citep{hc01,rpeb03}, variation among the mass distributions is  
one of the only distinguishing features among molecular cloud
populations.  Owing to relatively short molecular cloud lifetimes
\citep{bs80,mw-cluster-co,nanten-lifetime}, molecular clouds have
little time to increase significantly in mass due to cloud collisions
and accretion.  However, the destruction of molecular clouds by their
stellar progeny will change their mass through photodissociation and
hydrodynamic effects.  Observations show that the star formation rate
scales roughly with cloud mass in the Milky Way \citep{ms88}.  If this
is approximately correct throughout the Local Group, then differences
in the mass distribution of molecular clouds are not likely to arise
from different star formation rates.  It seems likely that differences
observed in the mass distributions must be due primarily to the
formation mechanism of molecular clouds.  Since many studies seek to
explain the mass distribution of molecular clouds
\citep{kwan79,fractal-mspec,vs97,stutzki98,wada00}, these
explanations must be expanded in scope to encompass the variety of
mass distributions observed in the Local Group.

It is interesting to note that the environment with the steepest index
of the mass distribution (M33) is also the region which is most
gravitationally stable with respect to gravitational instability
\citep{mk01}.  The behavior might be expected if two mechanisms
dominated the cloud formation process, each producing different mass
distributions and one of the mechanisms was regulated by gravitational
instability.  For example, if the molecular clouds that form in spiral
arms are more massive than those that form in the field, then a
steeper mass index is expected in M33 where the disk is stable.
Another possibility is that the galactic environment establishes the
cutoff mass for the mass distribution.  In both the inner Milky Way
and the LMC where there is reasonably clear evidence for a cutoff
mass, that mass is roughly $3\times 10^{6}~M_{\odot}$.  However, in
M33, the characteristic mass of molecular clouds must be smaller than
the completeness limit in the study ($1.5\times 10^{5}~M_{\odot}$) and
is likely $\sim 5\times 10^{4}~M_{\odot}$.  The outer Milky Way
does not appear to show a characteristic mass which can be attributed
to the absence of sufficient molecular material to populate the
distribution at masses near the cutoff.  It remains an open question
as to what physics would establish the characteristic mass in these
systems and why the characteristic mass in M33 would be two orders of
magnitude less massive than in the Milky Way and the LMC.

\section{Conclusions}

This study emphasizes the importance of performing a uniform analysis
to generate mass spectra.  Using the error-in-variables method of
parameter determination, we reanalyzed the molecular cloud catalogs
for the Local Group of galaxies and we report the following
conclusions:

1) Fits to the cumulative mass distribution using the
error-in-variables method produce a reliable estimate of the
parameters of the mass distribution.  Bootstrapping produces
reasonable uncertainties these parameters.  The adopted method is
superior to the standard technique of fitting a binned approximation
to the differential mass spectrum since it is insensitive to bin
selection and it also accounts for uncertainties in the mass estimate.

2) There is significant variation in the mass distributions of
molecular clouds across the Local Group even after accounting for
systematic effects and biases.  Differences in the method used to
catalog the molecular emission affect the derived parameters of the
mass distribution.  In particular, the presence and magnitude of a
cutoff in the mass distribution is affected by the decomposition
algorithm.  Unless the cutoff is quite significant (as it is in the
inner Milky Way), the presence of a truncation should be regarded with
some suspicion.  However, the index of the mass distribution is far
less sensitive to the particulars of the mass determination and
decomposition algorithm, resulting in systematic errors in the index
$\gamma$ of $\pm 0.1$. 

3) The mass distribution in the inner Milky Way has a measured index
of $\gamma=-1.5\pm 0.1$ with good evidence for a truncation in the
distribution setting a maximum mass of $10^{6.5}~M_{\odot}$.
Systematic errors particular to the study of the inner Milky Way
suggest that the true mass distribution may be steeper than this
derived value.  Using $\gamma\approx -1.5$ is appropriate for the
inner Milky Way but does not approximate the mass distribution of
molecular clouds across all galaxies.

4) The mass distribution of molecular clouds in the outer Milky Way is
significantly steeper than that found in the inner Galaxy.  The mass
distribution has an index of $\gamma=-2.1 \pm 0.2$, steeper than
previously claimed, and shows no evidence of a cutoff at high mass.

5) The GMCs in M33 show the steepest distribution found in this study
with no evidence of a cutoff.  It is possible that the distribution
actually has a cutoff below the completeness limit of the sample which
accounts for the derived index.

6) The LMC has a mass distribution that is steeper than that of the
inner Milky Way ($\gamma_{\mathrm{LMC}}=-1.7\pm 0.2$) but also shows
some evidence of a cutoff near $10^{6.5}~M_{\odot}$ which was unknown
heretofore.  An expanded catalog of clouds is needed to confirm this
result.

\acknowledgements  
This work is supported by an NSF postdoctoral fellowship
(AST-0502605).  I thank Adam Leroy for lengthy discussions regarding
parameter estimation and maximum likelihood.  I am grateful to Leo
Blitz for a careful reading of this work which, as always, improved
its clarity.  The comments of an anonymous referee helped to clarify
the motivation for this work.  This work relied heavily on the use of
NASA's Astrophysics Data System and the efforts of the Center for
Astrostatistics at the University of Pennsylvania.

\end{document}